\documentclass[%
reprint,
superscriptaddress,
altaffilletter,
amsmath, amssymb, aps, prl, floatfix
]{revtex4-1}

\usepackage{graphicx}
\usepackage{dcolumn}
\usepackage{bm}
\usepackage{comment}
\usepackage[usenames,dvipsnames]{color}
\usepackage{hyperref}
\hypersetup{colorlinks = true, allcolors = blue}
\usepackage[mathlines]{lineno}
\usepackage{multirow}
\usepackage[normalem]{ulem}

\DeclareGraphicsExtensions{.pdf,.png,.jpg}
\newcommand{\BBless}{\ensuremath{0\nu\beta\beta}}

\newcommand{\red}[1]{{\color{red}{#1}}}

\newcommand{\ckky}{counts/(keV$\cdot$kg$\cdot$yr)}
\newcommand{\ky}{kg$\cdot$yr}
\begin{document}

\newcommand{\CharacteristicROIEnergyResolutionBackground}{$(7.7 \pm 0.5)$~keV}
%


\newcommand{\TypicalRiseTimeMS}{100~ms}
\newcommand{\TypicalFallTimeMS}{400~ms}
\newcommand{\TypicalCalibrationEventRatemHz}{$\sim$50~mHz}
\newcommand{\TypicalPhysicsEventRatemHz}{$\sim$6~mHz}

\newcommand{\PercExposureHeaterTGS}{96.6}

\newcommand{\ExposureFirstDS}{37.6}
\newcommand{\ExposureSecondDS}{48.7}
\newcommand{\ResolutionFirstDS}{$8.3 \pm 0.4$}
\newcommand{\ResolutionSecondDS}{$7.4 \pm 0.7$}
\newcommand{\SigmaScaleFirstDS}{$0.95 \pm 0.07$}
\newcommand{\SigmaScaleSecondDS}{$1.01 \pm 0.06$}

\newcommand{\TotalSelectionEfficiencyFirstDS}{$85.7 \pm 3.4$} 
\newcommand{\TotalSelectionEfficiencySecondDS}{$94.0 \pm 2.9$}
\newcommand{\BaseSelectionEfficiencyFirstDS}{$95.63 \pm 0.01$} 
\newcommand{\BaseSelectionEfficiencySecondDS}{$96.69 \pm 0.01$}
\newcommand{\PSASelectionEfficiencyFirstDS}{$91.1 \pm 3.6$} 
\newcommand{\PSASelectionEfficiencySecondDS}{$98.2 \pm 3.0$}
\newcommand{\AccSelectionEfficiencyFirstDS}{$99.4 \pm 0.5$} 
\newcommand{\AccSelectionEfficiencySecondDS}{$100.0 \pm 0.4\ \,$}
\newcommand{\BackgroundIndxFirstDS}{$1.49 \pm 0.18$}
\newcommand{\BackgroundIndxSecondDS}{$1.35 \pm 0.19$}

\newcommand{\ExposureLossFromBadIntervals}{1\%}

\newcommand{\BBContainmentEfficiencyGFourWithError}{$88.35\pm 0.09$ }

%
%
\newcommand{\NumberOfCandidatesROI}{155}
\newcommand{\FinalTotalTeOExposure}{86.3~{\ky}}
\newcommand{\FinalCUOREIsotopeExposure}{24.0~\ky}

\newcommand{\BestFitZeroNuRateDefaultModel}{$(-1.0 ~^{+0.4}_{-0.3}~\mathrm{(stat.)}\pm 0.1~\mathrm{(syst.)})\times 10^{-25}$~yr$^{-1}$}
\newcommand{\CharacteristicROIBackgroundLevel}{$(0.014 \pm 0.002)$~\ckky}
%

%
%
\newcommand{\PercOfToysWithKSMetricWithinCL}{\red{XX}}
\newcommand{\KSMetricCL}{\red{XX}}

%
\newcommand{\PercProbForMostPositiveFluctuation}{\red{XX}}
\newcommand{\PercProbForMostNegativeFluctuation}{\red{XX}}
\newcommand{\PercProbForLargestFluctuationAnyWhereInROI}{\red{XX}}

%
%
\newcommand{\UpperLimitNuRateDefaultMethodStatOnly}{$\Gamma_{0\nu}<0.50\times 10^{-25}$~yr$^{-1}$}
\newcommand{\LowerLimitHalfLifeDefaultMethodStatOnly}{$T^{0\nu}_{1/2}> 1.4\times 10^{25}$~yr}
\newcommand{\MedianLowerLimitHalfLifeSensitivityDefaultMethod}{$7.0\times 10^{24}$~yr}
\newcommand{\PercProbToGetBetterLimitDefaultMethodStatOnly}{2}

\newcommand{\GlobalBoundOnEnergyScaleBias}{$\pm 0.5$~keV}

\newcommand{\BBlessShiftFromEnergyScale}{--}
\newcommand{\ScalingBBlessShiftFromEnergyScale}{0.2}
\newcommand{\BBlessShiftFromEnergyResolution}{--}
\newcommand{\ScalingBBlessShiftFromEnergyResolution}{1.5}
\newcommand{\BBlessShiftFromLineShape}{0.02}
\newcommand{\ScalingBBlessShiftFromLineShape}{2.4}
\newcommand{\BBlessShiftFromBkgShape}{0.05}
\newcommand{\ScalingBBlessShiftFromBkgShape}{0.8}
\newcommand{\BBlessShiftFromFitBias}{--}
\newcommand{\BBlessShiftFromFitBiasScaling}{0.3}
\newcommand{\BBlessSystErrorFromCutEfficiencies}{2.4\%}

\newcommand{\UpperLimitNuRateDefaultMethodWithSyst}{$\Gamma_{0\nu}<0.52\times 10^{-25}$~yr$^{-1}$}
\newcommand{\LowerLimitHalfLifeDefaultMethodWithSyst}{$T^{0\nu}_{1/2}> 1.3\times 10^{25}$~yr}
\newcommand{\CUOREHalflifeLimit}{\ensuremath{1.3\times10^{25}~{\rm yr}}}

\newcommand{\LowerLimitHalfLifeRolkeMethodWithSyst}{$T^{0\nu}_{1/2}> 2.1\times 10^{25}$~yr}
\newcommand{\MedianLowerLimitHalfLifeSensitivityRolkeMethod}{$7.6\times 10^{24}$~yr}

\newcommand{\UpperLimitNuRateCombinedAnalysisWithSyst}{$\Gamma_{0\nu}<0.47\times 10^{-25}$~yr$^{-1}$}
\newcommand{\UpperLimitNuRateCombinedRolkeAnalysisWithSyst}{$\Gamma_{0\nu}<0.31\times 10^{-25}$~yr$^{-1}$}
\newcommand{\LowerLimitHalflifeCombinedAnalysisWithSyst}{$T^{0\nu}_{1/2}> 1.5\times 10^{25}$~yr}
\newcommand{\LowerLimitHalflifeCombinedRolkeAnalysisWithSyst}{$T^{0\nu}_{1/2}> 2.2\times 10^{25}$~yr}

%
\newcommand{\AllCombinedHalflifeLimit}{\ensuremath{1.5\times10^{25}~{\rm yr}}}
%
\newcommand{\CombinedMbbLow}{110}
\newcommand{\CombinedMbbHigh}{520}

\preprint{APS/123-QED}
\title{First Results from CUORE:\\ A Search for Lepton Number Violation via $0\nu\beta\beta$ Decay of $^{130}$Te}
\author{C.~Alduino}
\affiliation{Department of Physics and Astronomy, University of South Carolina, Columbia, SC 29208, USA}


\author{K.~Alfonso}
\affiliation{Department of Physics and Astronomy, University of California, Los Angeles, CA 90095, USA}

\author{E.~Andreotti}
\altaffiliation{Present address: University College Leuven-Limburg, 3590 Diepenbeek, Belgium}
\affiliation{Dipartimento di Fisica e Matematica, Universit\`{a} dell'Insubria, Como I-22100, Italy}
\affiliation{INFN -- Sezione di Milano Bicocca, Milano I-20126, Italy}

\author{C.~Arnaboldi}
\affiliation{Dipartimento di Fisica, Universit\`{a} di Milano-Bicocca, Milano I-20126, Italy}

\author{F.~T.~Avignone~III}
\affiliation{Department of Physics and Astronomy, University of South Carolina, Columbia, SC 29208, USA}

\author{O.~Azzolini}
\affiliation{INFN -- Laboratori Nazionali di Legnaro, Legnaro (Padova) I-35020, Italy}


\author{I.~Bandac}
\affiliation{Department of Physics and Astronomy, University of South Carolina, Columbia, SC 29208, USA}

\author{T.~I.~Banks}
\affiliation{Department of Physics, University of California, Berkeley, CA 94720, USA}
\affiliation{Nuclear Science Division, Lawrence Berkeley National Laboratory, Berkeley, CA 94720, USA}

\author{G.~Bari}
\affiliation{INFN -- Sezione di Bologna, Bologna I-40127, Italy}

\author{M.~Barucci}
\altaffiliation{Present address: Istituto Nazionale di Ottica, Consiglio Nazionale delle Ricerche (INO-CNR), Firenze I-50125, Italy}
\affiliation{Dipartimento di Fisica, Universit\`{a} di Firenze, Firenze I-50125, Italy}
\affiliation{INFN -- Sezione di Firenze, Firenze I-50125, Italy}

% ...

\collaboration{CUORE Collaboration}\noaffiliation
\date{\today}
%
%
%
\begin{abstract}

The CUORE experiment, a ton-scale cryogenic bolometer array, recently began operation at the Laboratori Nazionali del Gran Sasso in Italy. The array represents a significant advancement in this technology, and in this work we apply it for the first time to a high-sensitivity search for a lepton-number--violating process: $^{130}$Te neutrinoless double-beta decay.  Examining a total TeO$_2$ exposure of {\FinalTotalTeOExposure}, 
characterized by an effective energy resolution of {\CharacteristicROIEnergyResolutionBackground}~FWHM and a background in the region of interest of {\CharacteristicROIBackgroundLevel}, we find no evidence for neutrinoless double-beta decay.  Including systematic uncertainties, we place a lower limit on the decay half-life of $T^{0\nu}_{1/2}(^{130}\mathrm{Te})>{\CUOREHalflifeLimit}$ (90\% C.L.); the median statistical sensitivity of this search is {\MedianLowerLimitHalfLifeSensitivityDefaultMethod}. Combining this result with those of two earlier experiments, Cuoricino and \mbox{CUORE-0}, we find $T^{0\nu}_{1/2}(^{130}\mathrm{Te})>{\AllCombinedHalflifeLimit}$ (90\% C.L.), which is the most stringent limit to date on this decay. Interpreting this result as a limit on the effective Majorana neutrino mass, we find $m_{\beta\beta}<({\CombinedMbbLow} - {\CombinedMbbHigh}$)~meV, where the range reflects the nuclear matrix element estimates employed.

\end{abstract}
\pacs{Valid PACS appear here}
\maketitle

The existence of nonzero neutrino masses is well established by precision measurements of neutrino flavor oscillation~\cite{PDG2016}. This discovery has given renewed impetus to long-standing questions as to the Dirac or Majorana nature of the neutrino~\cite{Majorana1937}, the role of Majorana neutrinos in cosmological evolution~\cite{FUKUGITA198645}, and the absolute neutrino mass. Neutrinoless double-beta ({\BBless}) decay is a lepton-number--violating process that can occur only if neutrinos are Majorana fermions~\cite{Racah1937,Furry1939,Pontecorvo:1967fh,BlackboxTheorem}. The discovery of this decay would unambiguously demonstrate that lepton number is not a symmetry of nature and that neutrinos are Majorana particles~\cite{NBBDReview2015}.
 
If it occurs, {\BBless} decay has a robust experimental signature: a peak in the summed energy spectrum of the final state electrons at the $Q$-value of the decay ($Q_{\beta\beta}$). To maximize sensitivity to this signature, an experiment must have a low background rate near $Q_{\beta\beta}$, good energy resolution, and a large source mass.  The Cryogenic Underground Observatory for Rare Events (CUORE)~\cite{CUOREREVIEW} is a new detector that applies the powerful macro-bolometer technique~\cite{Fiorini:1983yj,Enss:2008ek} at an unprecedented scale to search for {\BBless} decay of tellurium isotopes. In this work, we focus on {\BBless} decay of $^{130}$Te to the ground state of $^{130}$Xe. Our sensitivity benefits from the high natural abundance of $^{130}$Te, $(34.167 \pm 0.002)$\%~\cite{Fehr:2004jx}, and large $Q_{\beta\beta}$ of $(2527.515 \pm 0.013)$~keV~\cite{Redshaw:2009cf,Scielzo:2009co,Rahaman:2011wt}.  

CUORE is composed of 988 $5 \times 5 \times 5$ cm$^{3}$ TeO$_{2}$ crystals~\cite{Arnaboldi:2010fj}, each having a mass of 750 g, which we can cool to temperatures as low as 7~mK.   
When a crystal absorbs energy, we exploit the resulting temperature increase to measure that energy. Each crystal is instrumented with a thermistor~\cite{Haller:1984dr} to record thermal pulses, and a heater~\cite{Alessandrello:1998bf,Andreotti:2012zz} for thermal gain stabilization.

The crystals are arranged into 19 copper-framed towers, with each tower consisting of 13 floors of 4 crystals. The crystals are held in the tower frame by polytetrafluoroethylene supports. The towers are arranged in a close-packed array and thermally connected to the mixing chamber of a $^{3}$He/$^{4}$He dilution refrigerator\footnote{Leiden Cryogenics DRS-CF3000 continuous-cycle}, which is precooled by five two-stage ($\sim$40~K and $\sim$4~K) pulse tube cryocoolers~\footnote{Cryomech PT415-RM} and a Joule-Thomson expansion valve.

To suppress external $\gamma$-ray backgrounds, two lead shields are integrated into the cryogenic volume: a 30-cm thick shield at $\sim$50~mK above the detectors and a 6-cm thick shield at $\sim$4~K around and below the detectors. The lateral and lower shields are made from ancient Roman lead with extremely low levels of radioactivity~\cite{Alessandrello:1998163}. An external lead shield (25 cm thick) surrounded by borated polyethylene and boric acid (20 cm thick) provide additional shielding. More details on the experimental subsystems and shielding can be found in Refs.~\cite{CUORECryostat,CUORESuspension,RefFaraday,1748-0221-13-02-P02026,1748-0221-13-01-P01010}.

A prototype detector equivalent to a single CUORE tower, \mbox{CUORE-0}, operated
at Laboratori Nazionali del Gran Sasso from 2013 to 2015 and served to validate the 
materials and low-background assembly techniques used for CUORE~\cite{Alessandria:2012zp, CUOREAssemblyPaper, Alduino:2016vjd,Arnaboldi:2010fj,Andreotti:2009zza}. Before the current work, the strongest probe of $\beta\beta$~decay of $^{130}$Te came from \mbox{CUORE-0}~\cite{Alfonso:2015wka,Alduino:2016zrl,Alduino2017-CUORE02nu,JonsThesis}. 

The data presented here are from two month-long datasets collected from May to June (Dataset 1) and August to September (Dataset 2) of 2017. Between the two datasets, we improved the detector operating conditions; in particular, we implemented an active noise cancellation system on the cryocoolers~\cite{CUORE-PT-PhaseControl} and improved the electrical grounding of the experiment. The detector operating temperature is a compromise between minimizing the heat capacity of the crystals, thus maximizing the thermal gain, and optimizing the signal bandwidth. To select the optimal operating temperature, we performed a temperature scan to study the energy resolution achieved by a representative subset of detectors.  An operating temperature of approximately 15~mK was selected for both datasets.

Each dataset is bookended by periods devoted to energy calibration with $^{232}$Th $\gamma$-ray sources~\cite{Cushman:2016cnv}; the closing calibration is performed to verify the stability of the detector response over the dataset.
We use the data collected between calibrations, which we refer to as \emph{physics data}, for our $0\nu\beta\beta$ decay search.

The voltage across each thermistor is amplified and filtered~\cite{1748-0221-13-02-P02026,RefElectronics2,RefElectronics3,Arnaboldi:2010zz} and continuously digitized with a sampling rate of 1~kHz~\cite{AGiachero-Thesis,SDiDomizio-Thesis,SCopello-Thesis}.  A total of 984 of 988 channels are functioning. Thermal event pulses are identified by a software derivative trigger with channel-dependent thresholds ranging from 20 to a few hundred keV; we anticipate reducing these thresholds for future low-energy studies~\cite{DiDomizio:2011cv,Alduino:2017que}.  The rise and fall times of thermal pulses are on the order of {\TypicalRiseTimeMS} and {\TypicalFallTimeMS}, respectively.
We analyze a 10-s window consisting of 3~s before and 7~s after each trigger. The pre-trigger voltage provides a proxy for the bolometer temperature before the event, while we determine the event energy from the pulse amplitude.
The average event rate per detector is {\TypicalCalibrationEventRatemHz} in calibration data and {\TypicalPhysicsEventRatemHz} in physics data. In addition to triggered pulses, every few minutes each heater is injected with a stable voltage pulse ($\sim$1~ppm absolute stability)~\cite{1748-0221-13-02-P02029} to generate tagged reference events with fixed thermal energy. To monitor and characterize noise we also analyze waveforms
with no discernible thermal pulses.

To improve the signal-to-noise ratio we use an optimal filter~\cite{Gatti1986}, which exploits the distinct frequency characteristics of particle-induced and noise waveforms.  The pulse amplitude is determined from the maximum value attained by the filtered waveform. To monitor and correct for possible drifts in the energy-to-amplitude response of the detection chain (e.g., due to small drifts in operating temperature), which could otherwise spoil the energy resolution, we apply thermal gain stabilization (TGS) to each event amplitude. We apply one of two methods: the first uses monoenergetic heater pulses (heater-TGS), and the second uses pulses induced by $\gamma$ rays from the 2615-keV $^{208}$Tl calibration line (calibration-TGS).  Both methods were developed and used in \mbox{CUORE-0}~\cite{Alduino:2016zrl}. Heater-TGS is our default algorithm, while we use calibration-TGS for the $\sim$3\% of bolometers without functioning heaters and for channels in which calibration-TGS yields a statistically significant improvement in sensitivity compared to heater-TGS. In total, {\PercExposureHeaterTGS}\% of our exposure utilizes heater-TGS while the remainder uses calibration-TGS.

\begin{figure}[t]
\centering
\includegraphics[width=0.5\textwidth]{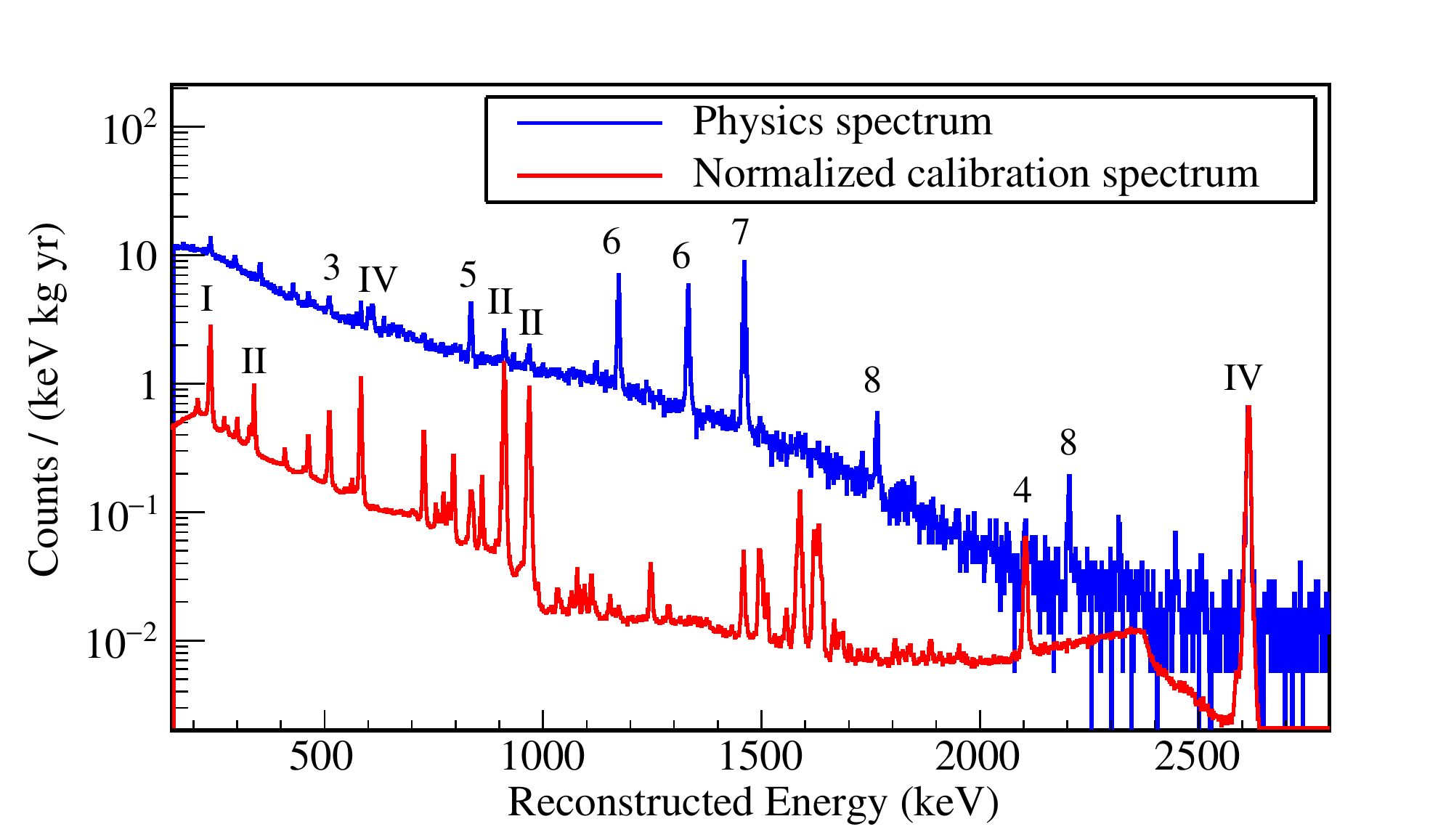}
\caption{Reconstructed energy spectra of physics (blue) and calibration (red) data. The calibration spectrum is normalized to the physics data at the 2615-keV line. The sources of the labeled peaks are identified as: (I)~$^{212}$Pb, (II)~$^{228}$Ac, (3)~$e^+e^−$ annihilation, (4/IV)~$^{208}$Tl, (5)~$^{54}$Mn, (6)~$^{60}$Co, (7)~$^{40}$K, (8)~$^{214}$Bi. Roman numbers indicate the spectral lines used for calibration.}
\label{fig:gamma_spectrum}
\end{figure}

To calibrate the detectors, we use six $\gamma$ lines from the $^{232}$Th calibration sources ranging from 239~keV to 2615~keV. We estimate the mean stabilized amplitude of each line and create a calibration function for each bolometer in each dataset (each {\it bolometer--dataset}), which maps stabilized pulse amplitudes to physical energies.  We find that the calibration functions of each bolometer--dataset are well described by a second-order polynomial with zero constant term throughout the calibrated energy range. After calibrating, to blind the region near $Q_{\beta\beta}$, we take events that reconstruct within 20 keV of the 2615 $^{208}$Tl line in physics data and move a blinded fraction of them down by 87~keV; this procedure produces an artificial peak at $Q_{\beta\beta}$~\cite{Alduino:2016zrl} and is later reversed once the {\BBless} search analysis is finalized.  The calibration and unblinded physics spectra are shown in \autoref{fig:gamma_spectrum}.

To select {\BBless} decay candidates in the physics data, we apply the following selection criteria. Firstly, we discard periods of noisy data caused, for example, by activity in the laboratory. This reduces the exposure by {\ExposureLossFromBadIntervals}.
Next, we impose basic pulse quality requirements to each event, requiring a single pulse-like feature in the event window and a stable pre-trigger voltage.
We then require the shape of each waveform to be consistent with that of a true signal-like event. We build a signal-like event sample in physics data from events that reconstruct within 10~keV of the $\gamma$ lines from $^{40}$K at 1461~keV and $^{60}$Co at 1173~keV and 1332~keV. We characterize event waveforms with six pulse-shape parameters and represent each event with a point in this 6-dimensional space. We calculate the Mahalanobis distance $D_{M}$~\cite{Mahalanobis:1936tj} for each event from the mean position of the signal sample. We choose the upper limit on $D_{M}$ that maximizes the discovery sensitivity~\cite{Cowan2011}.  Throughout this optimization, data from the region of interest for {\BBless} decay (ROI) are not used. In calculating the figure of merit for a given $D_{M}$ cutoff, we estimate the signal selection efficiency from $^{40}$K events near 1461~keV and the background selection efficiency from events with energy between $2700-3900$ keV. Events in this latter energy range are dominated by partially contained alpha particles and are representative of the dominant background in the ROI. Once the optimal $D_{M}$ cutoff is chosen, we evaluate the efficiency of the pulse shape selection using events belonging to the $^{208}$Tl 2615-keV line.

To reduce backgrounds from decays depositing energy in multiple crystals (e.g., $\alpha$ particles on crystal surfaces or multiple Compton scatters of $\gamma$ rays), we reject events that occur within 10~ms of an event in a different bolometer
(anti-coincidence selection). The width of the coincidence window is chosen after correcting for differences in detector rise times and trigger configurations that can affect the timestamp assigned to an event. The inter-bolometer timestamp differences are determined using physically coincident multi-detector events, such as pair-production events occurring in calibration data. The energy threshold for coincident events in the current analysis is set to 150~keV.
The anti-coincidence selection efficiency has two components: the probability for a {\BBless} decay to be fully contained in a single crystal and the probability to not accidentally coincide with another event. We estimate the former from simulation~\cite{Agostinelli:2003fg,Alduino:2017ods} and the latter we determine using the 1461-keV $\gamma$ ray from $^{40}$K electron capture, which is a single-event decay that is not expected to produce physical coincidences.

We evaluate the trigger efficiency as the fraction of tagged heater pulses that produce an event trigger. The heater pulse amplitude is scanned to study the energy dependence of the trigger efficiency. We also exploit heater events to measure the basic pulse quality selection efficiency mentioned above and the energy reconstruction efficiency (i.e., the probability that a monoenergetic pulse reconstructs correctly).
The combined trigger, basic pulse quality, and reconstruction efficiency, denoted by {\it base efficiency}, is averaged over all channels with functioning heaters and applied to all channels. In cases where a step in the event reconstruction procedure fails for a channel, we remove that channel from the subsequent analysis. The selection efficiencies are summarized in \autoref{tab:datasets}.
\begin{table}
\caption{Number of channels studied, event selection efficiencies, and performance parameters for the two datasets analyzed in this work. The effective resolution and background parameters are given at $Q_{\beta\beta}$. The uncertainty on exposure is negligible.}
\label{tab:datasets}
\rule{0pt}{4.5ex}
\begin{tabular}{l c c}\hline \hline
\rule{0pt}{2.5ex} &Dataset 1 &Dataset 2\\\cline{1-3}
\rule{0pt}{3.5ex}\textbf{Selection Efficiency\ (\%)} & & \\ \cline{1-1}
\rule{0pt}{2.5ex}Base  &{\BaseSelectionEfficiencyFirstDS} &{\BaseSelectionEfficiencySecondDS}\\
Pulse shape ($D_{M}$)&{\PSASelectionEfficiencyFirstDS} & {\PSASelectionEfficiencySecondDS}\\
Anti-coincidence (accidental) & {\AccSelectionEfficiencyFirstDS} & {\AccSelectionEfficiencySecondDS} \\
Anti-coincidence ($\beta\beta$ containment) & \multicolumn{2}{c}{\BBContainmentEfficiencyGFourWithError} \\
\rule{0pt}{2.5ex}Total (excl.\ $\beta\beta$ containment)  & {\TotalSelectionEfficiencyFirstDS} & {\TotalSelectionEfficiencySecondDS}\\
\cline{1-3}
\rule{0pt}{3.5ex}\textbf{Performance Parameters} & & \\\cline{1-1}
\rule{0pt}{2.5ex}Channels used & 876 & 935 \\
TeO$_2$ exposure (\ky) & \ExposureFirstDS & \ExposureSecondDS\\
Effective resolution (keV) & {\ResolutionFirstDS} & {\ResolutionSecondDS} \\
    Background ($10^{-2}$~c/(keV$\cdot$kg$\cdot$yr)) & {\BackgroundIndxFirstDS} & {\BackgroundIndxSecondDS}\\
\hline\hline 
  \end{tabular}
\end{table}

We establish the detector response to a monoenergetic event near $Q_{\beta\beta}$ using the high-statistics $^{208}$Tl 2615-keV $\gamma$ line from calibration data. The CUORE detectors exhibit a slightly non-Gaussian line shape, as was observed in \mbox{CUORE-0}~\cite{Alduino:2016zrl} and Cuoricino~\cite{Bryant:2010ua,Carrettoni:2011df}. The origin of this structure is under investigation; however, we model it empirically with a primary Gaussian component centered at 2615~keV and two additional Gaussian components, one on the right and one on the left of the main peak. We find this model provides a better description of the data compared to other models considered, for example, a single- or double-Gaussian photopeak. The choice of line shape is treated as a systematic uncertainty.
\begin{figure}
\centering
\includegraphics[width=0.5\textwidth]{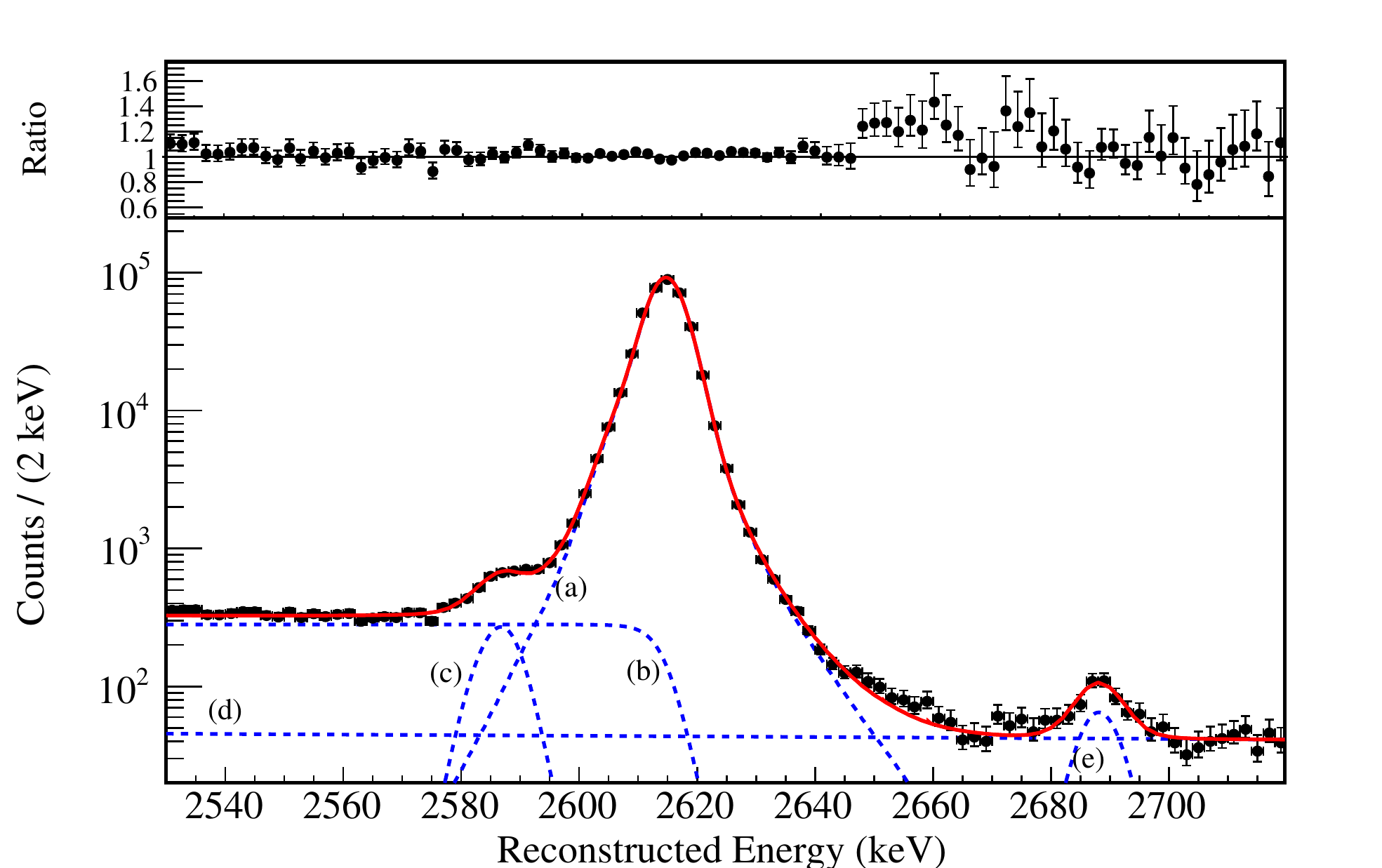}
\caption{Bottom: Sum of the results of the 19 tower-dependent UEML fits we use to estimate the line shape parameters of each bolometer--dataset in calibration data. The solid red line is the sum of the best-fit line shape model of each bolometer--dataset; the components of this summed best-fit model are shown by the blue dashed lines. We identify (a)~the multi-Gaussian photopeak that describes the detector response function, (b)~a multiscatter Compton contribution, (c)~multiple peaks due to 27--31~keV Te X-ray escape following an incident 2615-keV $\gamma$ ray, (d)~a linear continuum background due to coincident events, and (e)~a line due to coincident absorption of 2615-keV and 583-keV $\gamma$ rays from the $^{232}$Th decay chain followed by escape of a 511-keV annihilation $\gamma$ from pair production. Top: Ratio between calibration data and line shape model.}
\label{fig:lineshape}
\end{figure}
The three Gaussian components are parametrized with the same bolometer--dataset dependent width.  The normalized line shape function of each bolometer-dataset thus has 6 parameters: the means of the main peak and two subpeaks, the relative intensities of the subpeaks, and the common peak width.
We estimate the line shape parameters for each bolometer--dataset with a simultaneous, unbinned extended maximum likelihood~(UEML) fit performed on each tower in the energy range 2530--2720~keV. The simultaneous fit over a tower helps constrain common nuisance parameters such as relative intensity of x-ray escape peaks and continuum background. A simultaneous fit over the full array was not performed due to the computational demands.
A comparison of the fit results with the calibration data and a breakdown of the fit model are shown in \autoref{fig:lineshape}.

To characterize possible differences in the detector response between physics and calibration data we fit prominent background peaks in the physics data, with known energies between 800 and 2615 keV, using the best-fit line shape parameters determined above for each bolometer--dataset. At each energy this fit includes a dataset-dependent (i.e., channel independent) energy offset variable to parametrize energy misreconstruction. In addition, as the calibration line shape study was performed near 2615 keV, each fit includes a dataset-dependent(channel independent) energy resolution scaling variable to parameterize energy dependence of the resolution or a difference between background and calibration resolution. We find the energy misreconstruction is less than 0.5 keV over the calibrated energy range. The best-fit resolution scaling parameters at 2615 keV are {\SigmaScaleFirstDS} and {\SigmaScaleSecondDS} for the first and second dataset, respectively. To parametrize the energy dependence of the resolution scaling, we fit the set of scaling parameters determined at each peak energy studied with a quadratic function.  The resulting best-fit function is then used to estimate the resolution scaling at $Q_{\beta\beta}$.  The exposure-weighted harmonic mean energy resolution of the detectors (denoted {\it effective resolution}) in physics data, extrapolated to $Q_{\beta\beta}$, is given for each dataset in \autoref{tab:datasets}; to quote a single characteristic energy resolution for our entire exposure, we combine these, finding {\CharacteristicROIEnergyResolutionBackground} FWHM.

Before unblinding the physics data, we fix the model and fitting strategy to search for the {\BBless} decay of $^{130}$Te. The ROI is taken from 2465~keV to 2575~keV. The model for each bolometer--dataset is composed of a {\BBless} decay peak, a peak for $^{60}$Co coincident $\gamma$ rays, and a flat background. Each peak is modeled using the line shape discussed above, with the line width scaled by the resolution scaling extrapolated to the peak energy. All detectors are constrained to have the same {\BBless} decay rate $\Gamma_{0\nu}$, which we allow to vary freely in the fit; the position of the {\BBless} decay peak is fixed to $Q_{\beta\beta}$ for each bolometer-dataset.  
The $^{60}$Co peak position is a dataset-dependent free parameter; the $^{60}$Co rate is a single free parameter but the known isotope half-life is used to account for its decay.  The background rate is a dataset-dependent free parameter and is not scaled by the event selection efficiency.
\begin{figure}[]
\centering
\includegraphics[width=0.5\textwidth]{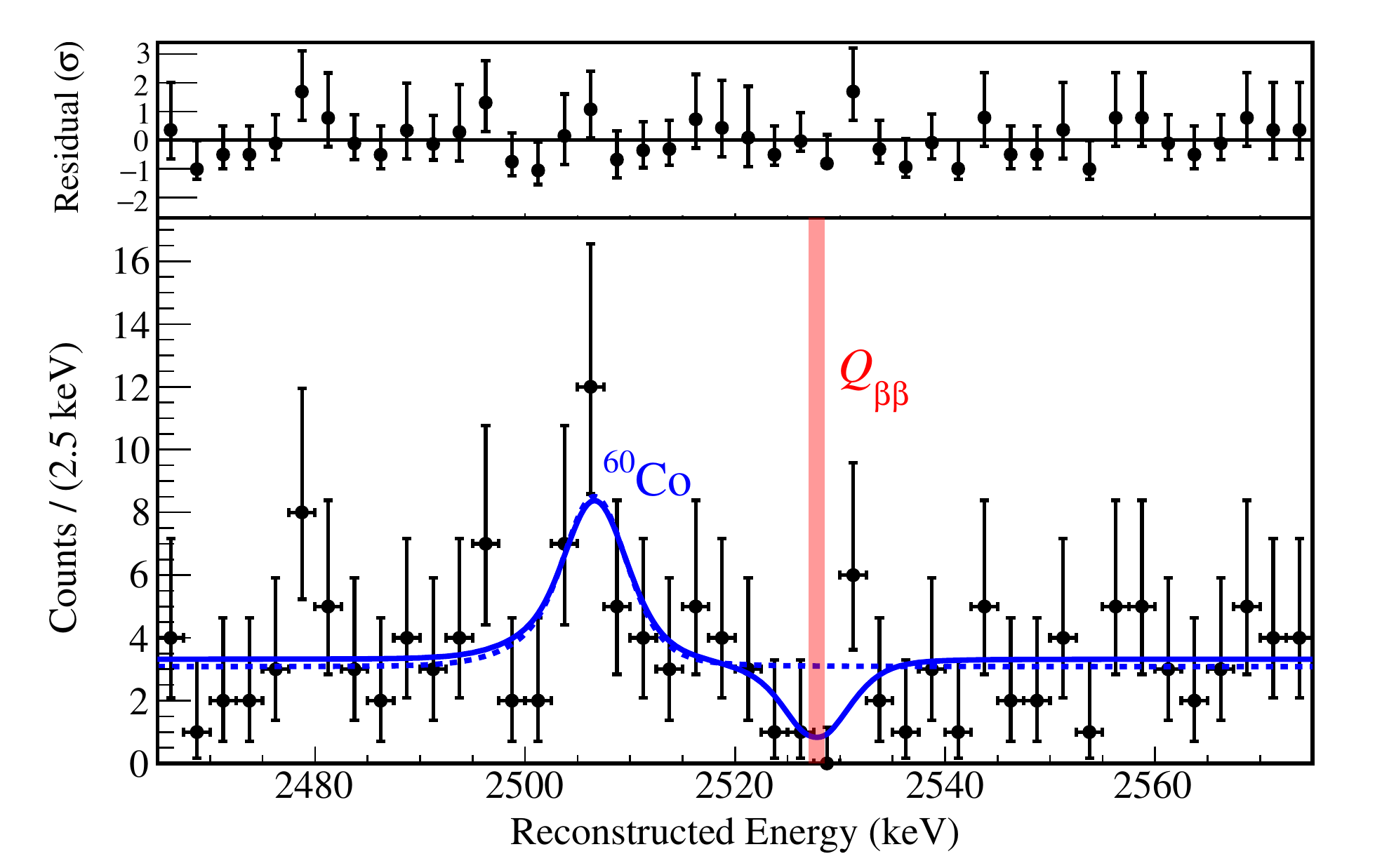}
\caption{Bottom: Best-fit model from the UEML fit (solid blue line) overlaid on the spectrum of $0\nu\beta\beta$ decay candidates observed in CUORE. The peak near 2506~keV is attributed to $^{60}$Co~\cite{Alduino:2016zrl}. The normalized residuals of this model and the binned data are shown in the top panel. The dashed (blue) curve shows the best-fit for a model with no {\BBless} decay component.  The vertical band is centered at $Q_{\beta\beta}$; the width of the band reflects the systematic uncertainty on the reconstructed energy.} 
\label{fig:unblinded_roi}
\end{figure}

\autoref{fig:unblinded_roi} shows the {\NumberOfCandidatesROI} candidate events in the ROI that pass all selection criteria together with the result of the UEML fit described above.  The total TeO$_2$ exposure is {\FinalTotalTeOExposure}, corresponding {\FinalCUOREIsotopeExposure} for $^{130}$Te. The best-fit $\Gamma_{0\nu}$ is {\BestFitZeroNuRateDefaultModel}.With zero signal, the best-fit background in the ROI averaged over both datasets is {\CharacteristicROIBackgroundLevel}.

To evaluate the goodness of fit, we prepare a large set of pseudo-experiments, each with a number of events determined by a Poisson distribution with a mean of {\NumberOfCandidatesROI} and energy distributed according to the best-fit zero-signal model. Repeating our {\BBless} decay search fit on each of these, we find that 68\% yield a negative log likelihood (NLL) larger than that obtained with our data.

We conclude there is no evidence for {\BBless} decay and set a 90\% confidence Bayesian upper limit on the rate, finding \mbox{{\UpperLimitNuRateDefaultMethodStatOnly}} (stat.~only) or {\LowerLimitHalfLifeDefaultMethodStatOnly}. In constructing the posterior pdf for $\Gamma_{0\nu}$, we approximate the marginalized likelihood with the profile likelihood and use a flat prior for $\Gamma_{0\nu}>=0$.  This approximation speeds up the computation and is valid when the marginalization is dominated by the most probable values of the nuisance parameters. We expect this for our likelihood as the number of events is large and the background dominates. To confirm this we perform an independent analysis using the BAT toolkit~\cite{Caldwell:2008fw} with the same prior but marginalize over the nuisance parameters. The results agree with those above to the percent level. 

We repeat our analysis on a large set of pseudo-experiments generated in the same way as for the goodness of fit study. We find the median 90\% confidence lower limit (sensitivity) for $T^{0\nu}_{1/2}$ is {\MedianLowerLimitHalfLifeSensitivityDefaultMethod}, and there is a {\PercProbToGetBetterLimitDefaultMethodStatOnly}\% probability of obtaining a more stringent limit than the one obtained with our data.

We estimate the systematic uncertainties following the same procedure used for \mbox{CUORE-0}~\cite{Alduino:2016zrl}. We perform a large number of pseudo-experiments with zero and nonzero signals assuming different detector line shape models and background shapes (flat and first-order polynomial), varying the energy resolution scaling parameters within their uncertainty, and shifting the position of $Q_{\beta\beta}$ by {\GlobalBoundOnEnergyScaleBias} to account for the energy reconstruction uncertainty. The results are summarized in \autoref{tab:systematics}. We find the fit bias on $\Gamma_{0\nu}$ to be negligible. Including these systematic uncertainties, the 90\% confidence limits are {\UpperLimitNuRateDefaultMethodWithSyst} and {\LowerLimitHalfLifeDefaultMethodWithSyst}. A frequentist analysis~\cite{Rolke:2004mj} yields {\LowerLimitHalfLifeRolkeMethodWithSyst} at 90\% C.L.\ with a median 90\% C.L.\ lower limit sensitivity for $T_{1/2}^{0\nu}$ of {\MedianLowerLimitHalfLifeSensitivityRolkeMethod}.

\begin{table}[]
\caption{Systematic uncertainties on $\Gamma_{0\nu}$ for zero signal (additive) and as a percentage of nonzero signal (scaling).}
\label{tab:systematics}
\begin{center}
  \begin{tabular}{l c c}
    \hline \hline
\rule{0pt}{2.5ex}    & Additive ($10^{-25}~\rm{yr^{-1}}$) & Scaling (\%)\\
    \hline
   Line shape & {\BBlessShiftFromLineShape} & {\ScalingBBlessShiftFromLineShape}\\
Energy resolution & {\BBlessShiftFromEnergyResolution} & {\ScalingBBlessShiftFromEnergyResolution}\\
Fit bias & {\BBlessShiftFromFitBias} & {\BBlessShiftFromFitBiasScaling} \\
Energy scale  & {\BBlessShiftFromEnergyScale} & {\ScalingBBlessShiftFromEnergyScale}\\
Background shape  & {\BBlessShiftFromBkgShape} & {\ScalingBBlessShiftFromBkgShape} \\\hline
\rule{0pt}{2.5ex}Selection efficiency & \multicolumn{2}{c}{\BBlessSystErrorFromCutEfficiencies}\\
\hline \hline
\end{tabular}
\end{center}
\end{table}

\begin{figure}[t!]
\centering
\includegraphics[width=0.5\textwidth]{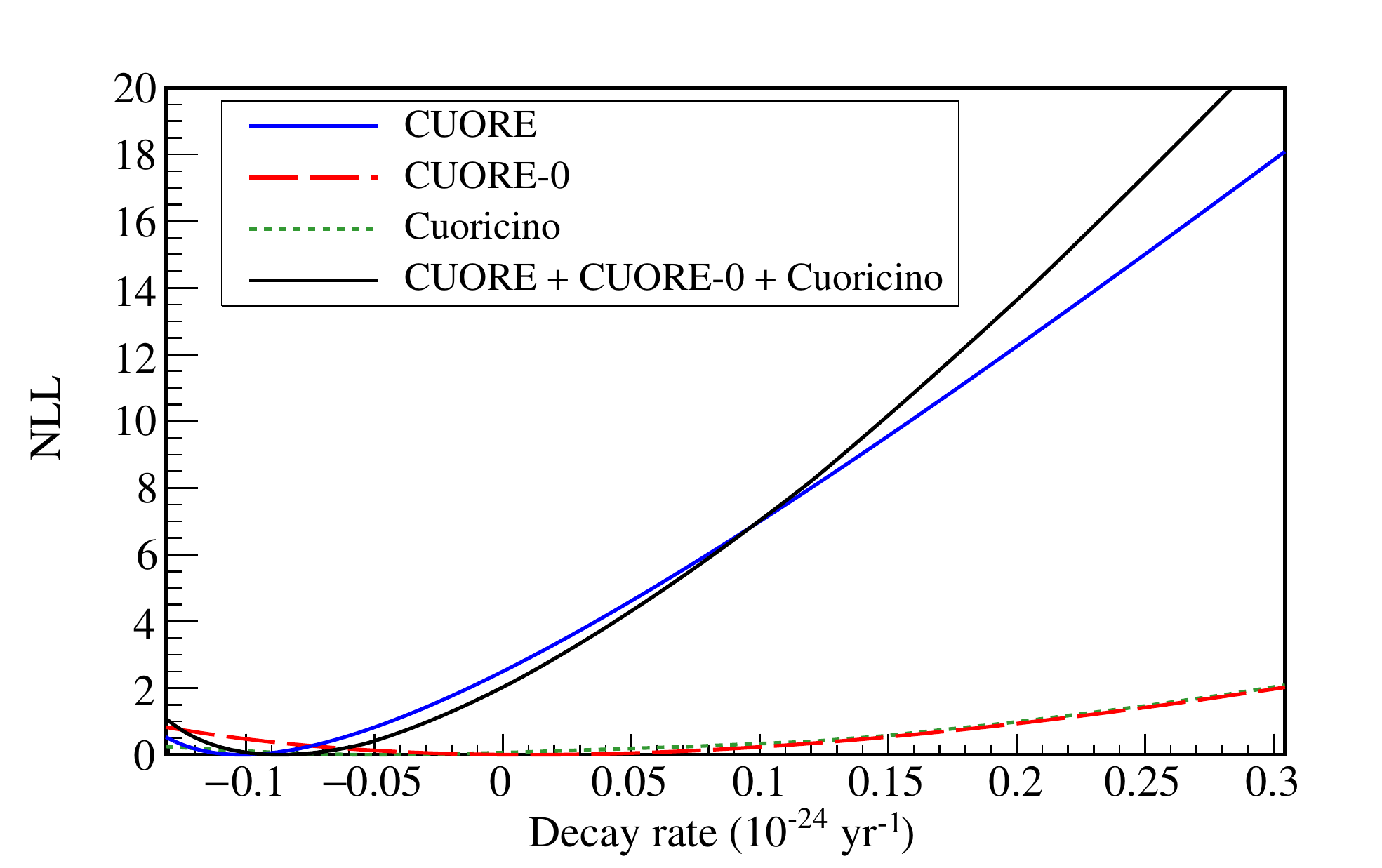}
\caption{Profile negative-log-likelihood curves for CUORE, \mbox{CUORE-0}, Cuoricino, and their combination.}
\label{fig:combined_nll}
\end{figure}

We combine our profile likelihood curve with those from 9.8~{\ky} of $^{130}$Te exposure from \mbox{CUORE-0}~\cite{Alfonso:2015wka} and 19.8~{\ky} from Cuoricino~\cite{Andreotti:2010vj} (see \autoref{fig:combined_nll}). The combined 90\% C.L.\ limits are {\UpperLimitNuRateCombinedAnalysisWithSyst} and \mbox{{\LowerLimitHalflifeCombinedAnalysisWithSyst}}. The frequentist technique yields \mbox{{\UpperLimitNuRateCombinedRolkeAnalysisWithSyst}} and {\LowerLimitHalflifeCombinedRolkeAnalysisWithSyst}.

We interpret the combined half-life limit, \mbox{{\LowerLimitHalflifeCombinedAnalysisWithSyst}}, as a limit on the effective Majorana neutrino mass ($m_{\beta\beta}$) in the framework of models of {\BBless} decay mediated by light Majorana neutrino exchange. We use phase-space factors from~\cite{Kotila:2012zza}, nuclear matrix elements from a broad range of models~\cite{Engel:2016xgb,Barea:2015kwa,Simkovic:2013qiy,Hyvarinen:2015bda,Menendez:2008jp,Rodriguez:2010mn,Vaquero:2014dna,PhysRevC.91.024316,Mustonen:2013zu,Neacsu:2014bia,Meroni:2012qf}, and  assume the axial coupling constant $g_A \simeq 1.27$;  this yields 
$m_{\beta\beta}<({\CombinedMbbLow} - {\CombinedMbbHigh}$)~meV at 90\% C.L., depending on the nuclear matrix element estimate employed.

In summary, we find no evidence for {\BBless} decay of $^{130}$Te and place the most stringent limit to date on this decay half-life.  The observed background, {\CharacteristicROIBackgroundLevel}, is in line with our expectations~\cite{Alduino:2017ods}. The characteristic energy resolution
at $Q_{\beta\beta}$ is {\CharacteristicROIEnergyResolutionBackground}, which we foresee improving to $\sim$5~keV by optimizing operating conditions and through analysis improvements. A study of our future sensitivity for a number of
scenarios is presented in Ref.~\cite{CUORE-Sensitivity}. The experimental progress in $0\nu\beta\beta$ decay searches has been dramatic in recent years; half-lives greater than $10^{25}$ yr are now probed by several experiments~\cite{PhysRevLett.117.082503,Albert:2014awa,Gerda:2017,Albert:2017owj}. CUORE is the first ton-scale cryogenic detector array in operation, more than an order of magnitude larger than its predecessors. The successful commissioning and operation of this large-mass, low-background, cryogenic bolometer array represents a major advancement in the application of this technique to {\BBless} decay searches.  


The CUORE Collaboration thanks the directors and staff of the Laboratori Nazionali del Gran Sasso and our technical staff for their valuable contribution to building and operating the detector. We thank Danielle Speller for contributions to detector calibration, data analysis, and discussion of the manuscript. We thank Giorgio Frossati for his crucial support in the commissioning of the dilution unit.
This work was supported by the Istituto Nazionale di
Fisica Nucleare (INFN); the National Science
Foundation under Grant Nos. NSF-PHY-0605119, NSF-PHY-0500337,
NSF-PHY-0855314, NSF-PHY-0902171, NSF-PHY-0969852, NSF-PHY-1307204, and NSF-PHY-1404205; the Alfred
P. Sloan Foundation; and Yale
University. This material is also based upon work supported  
by the US Department of Energy (DOE) Office of Science under Contract Nos. DE-AC02-05CH11231 and
DE-AC52-07NA27344; and by the DOE Office of Science, Office of Nuclear Physics under Contract Nos. DE-FG02-08ER41551, DE-FG03-00ER41138, DE-SC0011091, and DE-SC0012654. 
This research used resources of the National Energy Research Scientific Computing Center (NERSC).

\bibliography{bibliography}
\end{document}